
\magnification=\magstep1
\def\et{{\sl et al}~}
\def\ltorder{\mathrel{\raise.3ex\hbox{$<$}\mkern-14mu
             \lower0.6ex\hbox{$\sim$}}}
\font\BF=cmbx10 scaled \magstep1
\bigskip
\centerline{\BF STATISTICS OF N-BODY SIMULATIONS.}
\centerline{\BF II. EQUAL MASSES AFTER CORE COLLAPSE}
\bigskip
\centerline{Mirek Giersz\footnote{$^1$}{On
leave from N.  Copernicus Astronomical Centre, Bartycka 12, 00-716 Warsaw,
Poland} and Douglas C.
Heggie}
\medskip
\centerline{University of Edinburgh,}
\centerline{Department of Mathematics and Statistics,}
\centerline{King's Buildings,}
\centerline{Edinburgh EH9 3JZ,}
\centerline{U.K.}
\bigskip
\vfill\eject
{\bf ABSTRACT}
\medskip

This paper presents and analyses statistical results from a large number
of $N$-body simulations of isolated systems with equal masses, in which
$250\le N\le 2000$.  It concentrates on the phase starting around the
end of core collapse.  Binaries play a crucial role, and we find that
the total energy of bound pairs is in line with theoretical
expectations.  Interpretation of the total {\sl number} is complicated
by the presence of a number of binaries on the hard/soft threshold.
Interactions of hard binaries are consistent with the Spitzer (1987)
cross section.

The spatial evolution of the half-mass radius after core collapse nearly
follows classical theory, and, by comparison with Fokker-Planck and gas
models, allows a redetermination of the effective thermal conductivity
and the argument of the Coulomb logarithm in the expression for the
relaxation time.  The evolution of the inner parts of the system around
the time of core bounce is consistent with these simplified models
provided that the continuous production of energy, as usually assumed,
is replaced by a model of {\sl stochastic} energy production.
Similarly, post-collapse evolution of the core requires a modest
recalibration of the coefficient of energy generation, especially for
small $N$.  These remarks refer to the behaviour averaged over many
models; individual cases show alternate and irregular phases of
expansion and recollapse.  The distributions of velocity dispersion and
anisotropy become remarkably homologous soon after core bounce.

The bound mass of the systems very nearly follows a power law with time.
A small number of escapers, presumed to be those associated with binary
activity, dominate the energy which is carried off: the distribution of
energies of escapers changes abruptly at the end of core collapse.  The
``internal" energy of escaping binaries is consistent with theoretical
expectations, and again supports Spitzer's reaction cross section for
hard binaries.

\bigskip


\bigskip
\vfill\eject
{\bf 1.  INTRODUCTION}
\medskip
{\bf 1.1  The Statistical Study of Idealised $N$-body Models}
\smallskip
The ultimate purpose of this series of papers is the improved
understanding of the evolution of star clusters.  In this context the
study of isolated $N$-body systems in which all stars have the same mass
may seem unrealistic.  And yet the dynamical study of real
star clusters leans heavily on the theory of several complicated
processes, such as the evolution
of binary stars, stellar escape, the generation of anisotropy, and so
on.  What we show in this series of papers is that a careful study
of many relatively small and idealised $N$-body models allows one to test
and, in some cases, improve this theory.

Our strategic approach to these questions was discussed in Paper I
(Giersz \& Heggie 1994).  There we argued that the combined study of large
numbers of $N$-body models, which may be done rather easily with parallel
computers, leads to results of sufficiently high statistical quality
that several old problems of stellar dynamics can be reexamined
empirically, leading in turn to new developments in theory.  An example
from the previous paper is the study of the escape rate, which showed
the crucial role of the development of anisotropy, as well as that of the
spatial evolution of the system.

Paper I described results up to a point late in core collapse, and the
present paper takes up the analysis of our data at that point.  One
reason for dividing the discussion there is that many aspects of the
subsequent evolution of the systems are heavily influenced by the
formation and evolution of binary stars.  This topic takes up Section 2,
and there follow two sections dealing, respectively, with the evolution
of the entire spatial structure and the statistical properties of the
velocities of the stars.  Sections 5 and 6 then discuss in some detail
conditions in the core of the systems, and aspects of the escape of
stars and binaries, respectively.
The final section sums up.

Further papers in this series will extend our approach towards
progressively more realistic models for star clusters.  Part of our
purpose here is to examine the conclusions of the survey, using the
isotropic Fokker-Planck model, by Chernoff \&
Weinberg (1990).  It has proved to be one of the most valuable resources
for understanding the evolution of globular clusters.  Thus our next paper
discusses the effects of a spectrum of stellar masses similar to that
chosen by them.  Later papers consider the influence of tidal effects
and stellar evolution (modeled as instantaneous mass loss).

\medskip
{\bf 1.2  The Models}
\smallskip

In this paper we make use of the same sets of models, with $N = 250$, $500$,
$1000$ and $2000$,
as in Paper I, which gives all necessary details of initial
conditions, hardware and software considerations, and the kind of data
which is available for study. We merely recapitulate the units used, in
which the initial total mass and energy are $M = 1$ and $E = -1/4$,
respectively, and the constant of gravitation $G = 1$.  We refer to these
hereafter as ``$N$-body units''.
There is, however, one issue which was
irrelevant for Paper I, which is the completeness of the samples.

It was our intention to study models for a time comparable with at least
two core collapse times, which implies a time approximately proportional
to $N$. Because the computational effort grows rapidly with $N$, a
compromise was reached in which we attempted to integrate models up to a time
$t_{end}$ given (in $N$-body units) in Table 1.
Even so, some models were lost
before time $t_{end}$ because of hardware or software problems, and so
for
each series Table 1 shows how many
models were running at the final time.  Even those models which have
been successfully completed may be at rather different evolutionary
stages at the finishing time:
different $N$-body models, drawn from the same initial conditions except
for the choice of random numbers, evolve at surprisingly different
rates.

There is little doubt that the time at which some models are lost
(because of software problems) is closely linked with their dynamical
behaviour, e.g.  the prevalence of time-consuming few-body interactions.
Coupled with the substantial spread of evolution rates, this can in
principle seriously bias the statistical results, and should strictly
be avoided.  In some of our results, however, we do display data
extending to a time at which a few of the models have already stopped,
and the reader should bear Table 1 in mind when conclusions are drawn
from our results.

\bigskip

{\bf 2.  BINARIES}
\medskip
Many aspects of the evolution of $N$-body systems, from around the time of core
bounce, are dominated by binaries.  Examples include the radius of the
core and the energy of escapers.  One of our aims in this paper is to
characterise the evolution of the binaries in terms of relatively simple
models, so that the application of these models in studies of larger
systems can be carried out with greater confidence.  There are three
broad classes of models: (i) simple scaling arguments, as in \S2.1; (ii)
simple models for binary evolution, such as those which are often
incorporated into
Fokker-Planck and gas models (cf. eq.(13)); and (iii) rather detailed
models on the stochastic evolution of individual binaries, based on
appropriate reaction cross sections, as in \S6.2.  The third group of
models can also be incorporated into Fokker-Planck and gas
models for the evolution of the cluster (cf.\S3.2).  Most of these
models can be tested at various levels, including (i) statistics on the
behaviour of
individual binaries (e.g. \S2.4); (ii) statistics on their numbers and
energy (e.g.\S\S2.2, 2.3), and (iii)
the effect they have on the structure of the cluster (e.g. \S3.2).
These remarks should give the reader a point of reference when some of
the numerous combinations of model and test data are discussed in this
and succeeding sections.

\medskip
{\bf 2.1  Time of First Formation of Binaries}
\smallskip
At one time it was thought that the role of
binary stars becomes relatively less important the larger the value of
$N$ (Spitzer \& Hart 1971).  Now it is accepted that, at least in the
point-mass approximation, binaries play a crucial role in the
post-collapse evolution
of systems of all sizes (e.g. Spitzer 1987).  However, provided that
there are no primordial binaries, binaries play a relatively minor role
throughout most of core collapse, especially for large $N$.  The purpose
of this section is to quantify this statement on the basis of $N$-body
models with $250 \le N\le 2000$.

First we outline a theoretical argument (cf. Inagaki 1984).  The rate of
formation of hard
binaries by 3-body encounters in a system with $N$ stars varies as
$$
\dot N_{b} \sim {1\over {N t_r}}, \eqno(1)
$$
where $t_r$ is the relaxation time.  We shall apply
this to the core (replacing $N$ by $N_c$ and $t_r$ by $t_{rc}$ to
indicate values appropriate to the core), and assume that $N_c$ varies
as a power of $t_{rc}$ during late core collapse.  If we set
$$
{N_c\over N_c(0)} = \left(t_{rc}\over
t_{rc}(0)\right)^{\displaystyle{{2(3-\alpha)\over 6-\alpha}}}, \eqno(2)
$$
where a zero denotes an initial value, and we would have $\alpha = 2.21$ in
homologous core collapse in a gaseous model (Lynden-Bell \& Eggleton 1980),
it follows that
$$
\dot N_b\sim {1\over t_{rc}(0) N_c(0)}\left({t_{rc}\over
t_{rc}(0)}\right)^{\displaystyle{{-{3(4-\alpha)\over 6-\alpha}}}}. \eqno(3)
$$
Then, since $t_{rc}$ varies linearly with time in homologous core
collapse, and $N_b(0) = 0$, we may estimate that
$$
N_b \sim {1\over N_c(0)}\left({t_{rc}\over
t_{rc}(0)}\right)^{\displaystyle{-{2(3-\alpha)\over 6-\alpha}}} \eqno(4)
$$
by the time $t_{rc}\ll t_{rc}(0)$.
Therefore the first binary should form when
$$
{t_{rc}\over t_{rc}(0)}\sim N_c(0)^{\displaystyle{-{6-\alpha\over
2(3-\alpha)}}}. \eqno(5)
$$

Within our approximations this can be read as the fraction of the
collapse time which remains when the first binary forms.  It is a steep
power of $N_c(0) \propto N$, varying as $N^{-2.4}$ approximately for
homologous collapse.  Therefore the larger the system the more abrupt is
the appearance of the first hard binary late in core collapse.

This is a statistical result, and we have found that individual $N$-body
systems vary widely, especially when $N$ is not very large.  The
empirical results also depend on the definition adopted for a ``hard"
binary.  If $3kT/2$ is the initial mean kinetic energy of single stars,
binaries with an energy exceeding $1kT$ are present in a fraction of
systems initially when $N\le500$, and for $N=250$ a few systems
initially have
binaries with energies exceeding $3kT$.  Table 2 indicates, for each $N$
and for three different binding energies the times at which half our
systems contained a binary with energy above the threshold.  It is
evident that our largest and smallest systems (with $N = 2000$ and $250$,
respectively) represent opposite extremes of behaviour.  For $N = 250$
the formation and evolution of binaries proceeds for much of the core
collapse phase, whereas for $N = 2000$ they are present only during
the last 10\% or so of its duration.  Indeed this result implies that
our estimate of the
$N$-dependence in eq.(5) cannot
apply for values of $N\ltorder 250$.

In principle a more detailed comparison between theory and $N$-body
data can be carried out by computing the rate of formation of binaries
in an evolving gas or Fokker-Planck model.  For this purpose we chose
values of the free parameters in these models (a coefficient in the
Coulomb logarithm and, for the gas model, a coefficient in the thermal
conductivity) as in Paper I.  For the rate of formation of binaries we
used the result of Hut (1985).  The comparison is not at all
straightforward, as Hut's formula is intended to yield the rate of
formation of ``permanent" hard binaries, i.e. those which are not
subsequently disrupted in encounters (though they may be ejected from
the system).  This is difficult to find for our $N$-body models.
Goodman
\& Hut (1993) found that the median energy of a newly formed permanent
binary is $2.9kT$.  Even so, we have found that the theoretical time
of formation of the first permanent hard binary (thus defined) is no
earlier than the time $t_{10}$ in Table 1, and indeed still later for
the larger values of $N$.  It is possible that the time-averaged rate
of formation of binaries is larger than predicted theoretically
because of fluctuations in the density of the core which are not
modelled in the continuum models.

\medskip
{\bf 2.2   Number of Binaries}
\smallskip

We now discuss one aspect of the post-collapse phase of the evolution,
for the first time in these papers.  Following H\'enon (1965), theory
requires that the overall expansion of the cluster is powered by the
evolution of binaries.  Since the expansion takes place on the time
scale of the half-mass relaxation time $t_{rh}$, the required power
can be estimated as
$$
\dot E\sim {\vert E\vert\over t_{rh}}, \eqno(6)
$$
where $E$ is the total ``external"
energy of the system, i.e. excluding the internal binding energy of
binaries. The energy yielded by each binary may be assumed
to be proportional to the central potential $\phi_c$ (Goodman 1984, 1987,
Statler \et 1987), and so the formation rate must vary as $\dot
N_b\propto E/(m\phi_ct_{rh})$, where $m$ is an individual stellar mass.
Assuming that the part of the cluster
between the core- and half-mass radii is nearly isothermal, it follows that
$$
\dot N_b \propto {N\sigma_c^2\over \phi_c t_{rh}},\eqno(7)
$$
where $\sigma_c$ is the
root mean square one-dimensional velocity dispersion in the core.  Applying
eq.(1) to the
core we
we deduce that
$$
N_c\propto \left({N\vert\phi_c\vert\over \sigma_c^2}\right)^{1/3}, \eqno(8)
$$
which we discuss in \S5.2.

This argument about the formation rate says nothing about the average
numbers of binaries, which depends also on their individual lifetime
within the cluster, $t_b$ say.  If each binary stayed within the core
until ejection we could estimate that its hardening rate is
$\sim m\sigma_c^2/t_{rc}$ (Heggie 1975, Hills
1975, though this formula is valid only if we neglect variations in
the Coulomb logarithm).  Hence we could estimate
that
$$
t_b\propto t_{rc}{\left (\vert\phi_c\vert\over \sigma_c^2\right)}. \eqno(9)
$$
If we do not assume
that binaries remain within the core then we can still express our
result as
$$\eqalignno{
N_b&\sim \dot N_b t_b&\cr
&\propto \left( {t_b\sigma_c^2\over \vert\phi_c\vert
t_{rc}}\right)\left( {t_{rc}\over t_{rh}}\right) N &(10)\cr
}$$
by eqs.(7) and (9).
Using the isothermal approximation again, and eq.(8), it follows that
$$
N_b\propto
\left({t_b\sigma_c^2\over{\vert\phi_c\vert t_{rc}}}\right)
\left({\vert\phi_c\vert\over\sigma_c^2}\right)^{2/3}N^{-1/3}. \eqno(11)
$$

The empirical data from our $N$-body models is somewhat at variance with
the simplest interpretation of this formula, i.e. $N_b\propto N^{-1/3}$,
since we find in the post-collapse regime that the average number of
binaries (defined here as regularised binaries (Aarseth 1985)) actually
tends to {\sl increase} with $N$, but quite weakly, and with large
fluctuations (Fig.1).  The $N$-dependence is roughly $N_b \propto N^{0.2}$.
A simple possible explanation of this discrepancy, and one for which
evidence is  presented in the next section, is that
our count of binaries includes temporary pairs which contribute
nothing to the energetics of the expansion, but whose numbers disguise
the smaller number of active binaries.  It is possible, however, that
$t_b$ much exceeds the estimate given by eq.(9).  As a binary hardens it
tends to be ejected ever further from the core (Hut \et 1992), and dynamical
friction acting on a time scale more comparable with $t_{rh}$ than with
$t_{rc}$ may determine its lifetime.  This would lead roughly to $N_b\propto
N$, which differs from the empirical data in the opposite sense but can be
ruled out even more strongly.  It is quite possible, therefore, that the
data of Fig.1 are explicable if the regularised binaries consist of some
temporary binaries with lifetimes (to destruction) of order $t_{rc}$, and
others whose lifetimes (to ejection) are of order $t_{rh}$.

These theoretical comparisons are based on very simple scaling arguments.
It is also possible to compute the numbers of binaries that would be
formed in gas- and Fokker-Planck models of star clusters.  We have
modeled the formation and evolution
of binaries in these models in several different
ways (cf. \S3.2 below), and all predict that the formation of binaries
during late core collapse occurs too slowly in comparison with $N$-body
data (cf. \S2.1).  Here, however, we are concerned with the predicted
numbers after core collapse.

The standard way of treating these processes in gas and Fokker-Planck
models is to assume that binaries form at a smooth rate given by a
refined and local version of eq.(1), and that they instantly emit an
amount of energy of order $\vert\phi_c\vert$.  Therefore these models
cannot be used to predict the current number of binaries bound to the
system. The simplest basis for comparison with the $N$-body data is to
compute the total number of binaries formed in the continuum model and
to compare this with $N$-body data on the sum of all regularised bound
binaries plus all regularised binaries which have been removed from the
system.  Even so, the predicted numbers of binaries are smaller than
those obtained from the $N$-body models, and the discrepancy is bigger
for larger $N$.

Gas models which treat the formation of binaries and their subsequent
burning as a {\sl stochastic} process can reproduce a nearly constant
number of bound binaries after core collapse, as found in $N$-body
models (see Fig.1). For different models the number levels off at
different values, which depend on the assumptions about how much of
the energy generated by binaries is directly supplied to the core, and
on the form of the cross section for binary hardening (cf. \S 2.4
below). We did not, however, attempt to evaluate the stochastic gas
models by comparing the empirical and predicted numbers of binaries,
because the empirical ($N$-body) data are contaminated in at least in
two ways. First, as was mentioned in \S2.1, it is very difficult to
distinguish between permanent and non permanent binaries for $N$-body
models. Second, hard binaries can spend a substantial time in the
halo, because interactions with field stars remove them from the core,
and that process was not simulated in our gas models. Therefore we can
expect that the ``real'' number of binaries which contribute to the
cluster energy balance (which is appropriate for comparison with the
gas models) should be smaller (by a factor which may be of order
unity) than the value obtained from $N$-body data. The $N$-dependence
of the number of binaries obtained from stochastic gas models is
similar to that predicted by eq.(11), (with $t_b$ estimated as in
eq.(9)), in the sense that the number of binaries is smaller for
larger systems, but the power-law index is slightly smaller. The
average value over all models is about $-0.13$, whereas the predicted
value is about $-0.2$ when we include the observed $N$-dependence of
the scaled central potential $\phi_c/\sigma_c^2$ (cf. \S5.1 and
Fig.9).

A hybrid way of testing the theoretical binary formation rate is to
take the $N$-body data on the spatial structure and the velocity
dispersion profile, and to apply to it the standard formula (Hut 1985)
for the local rate of formation of permanent hard pairs.  Just as for
the theoretical production of hard binaries in gas models (\S2.1),
this predicts a population of binaries that increases too slowly in the
collapse phase. In the post-collapse phase, however, it produces generally
(except for $N=250$) somewhat too few binaries (i.e. by comparison with the
total number of bound and unbound regularised binaries
in the $N$-body data). A very clear trend with $N$ can be observed, in
the sense that the
discrepancy between the actual and predicted numbers of binaries increases
with $N$. This trend can be at least partially explained by the way in which
the predicted number of binaries is estimated from the $N$-body data. The
rate of binary formation is a strong function of density, which in turn is
mainly determined by the innermost Lagrangian radii. For decreasing $N$ the
number of stars
in each Lagrangian shell decreases, and so the density estimation is poorer
and has larger fluctuations. Positive density fluctuations bias the
predicted number of binaries to high values, and this bias is greater
for smaller systems.

Most of the remarks in this section refer to the early post-collapse phase.
For systems in which we can observe the long term post-collapse evolution
($N = 250$ and $500$) the estimated and computed rates of
binary formation become very similar. In a sense this finding supports
H\`enon's theory
that, in post-collapse evolution, energy generated by binaries
is adjusted to the energy demanded by the overall expansion. It appears that
most of the discrepancies between the theoretical and $N$-body
models arise near core bounce.

\medskip
{\bf 2.3 Energy of Binaries}
\smallskip

We now consider the internal energies of hard binaries which are bound
to the system.  The lower limit for an individual binary is set by the
definition of `hard', which is often taken to mean energies above $1kT$.
The upper limit is set by the escape of binaries following energetic
interactions: the mean energy imparted to the barycentric motion of the
binary is a certain fraction of its internal energy, and so the binary
is likely to escape after an energetic interaction if its internal
binding energy exceeds a certain multiple of the escape energy.

Because of the stochastic aspects of such interactions, it is not
surprising that there are wide variations in the maximum binding
energy of binaries in our $N$-body models; the maximum occasionally
reaches values comparable with the entire binding energy of the
system.  The {\sl average} (over all cases at a given time) of the
maximum individual binding energy of a binary is quite well behaved,
however (Fig.2).  Models for all $N$ show a slow decline during the
post-collapse phase, and this decline is consistent with supposing
that the maximum is approximately proportional to the escape energy
from the centre (Fig.9 and \S6.3).

When we turn to the {\sl total} internal energy of bound binaries
(which we refer to as $E_{b,int}$) there is a trend with $N$ which is
more consistent with theoretical expectations than is the case for the
observed {\sl number} of binaries.  At comparable times in the
post-collapse phase the total energy is smaller for larger systems.
The maximum occurs early in the post-collapse phase, and may be
estimated very roughly as $E_{b,int} \simeq 0.06(N/1000)^{-1/3}$; the
power is suggested by the arguments of \S2.2 (i.e. eq.(11), and
assuming that the energy of each binary is proportional to
$\sigma_c^2$ or $\phi_c$, and that $t_b\sim t_{rc}$), and the
coefficient is determined from the $N$-body data. The fact that the
$N$-dependence is roughly consistent suggests that the total energy is
almost entirely contributed by at most a single hard binary, whereas
the balance of the observed numbers of binaries is made up by much
less energetic pairs which are almost always present after core
collapse.

Besides this roughly quantified $N$-dependence at comparable times,
there is a noticeable decline with time in the total energy of binaries.
This nearly follows a similar trend, noted above, in the maximum energy
of all binaries.

It was noted in \S2.1 that the first hard binaries tend to form more
quickly than one would expect in comparison with theoretical results on
the formation rate of ``permanent'' hard pairs.  The opposite effect
occurs (i.e. binaries are less effective than expected) if one computes the
energy generated by binaries. In a gas model this is often estimated using the
following rather standard formula for the rate of generation of energy per unit
mass:
$$
\varepsilon = 7.5G^5f_2f_3m^3
\left({\vert\phi_c\vert\over\sigma_c^2}\right)
\rho^2 \sigma^{-7}, \eqno(12)
$$
(cf. Goodman 1987), where $m$ is the mass of an individual
star, $\rho$ is the density, $\sigma$ the one-dimensional velocity dispersion,
and $f_2$ and $f_3$ are coefficients
(of order unity) which determine the rate of binary formation and how much
energy released by binaries goes directly to the core, respectively.
In later discussion we shall sometimes refer to the entire
dimensionless part of the coefficient as $C_b$, i.e.
$$
\varepsilon = C_bG^5m^3\rho^2\sigma^{-7}.\eqno(13)
$$
A value of $C_b = 90$ is often used (Cohn 1985).
In our $N$-body systems
the energy in hard binaries catches up with the prediction of this
formula late in core collapse, but up until then the predicted energy
generation is too large.  We consider that the cause of this discrepancy
is the assumption, implicit in the formula, of ``instantaneous burning",
i.e. that a newly formed binary will instantly yield the energy which,
in fact, is spread out over its lifetime.  This issue is discussed
further in \S3.2 below. On the other hand, during
the late pre- and early post-collapse phase
(around the time of core bounce) the energy generated in $N$-body systems is
much bigger than in the gas models. In other words, the $N$-body models
appear to require energy to be supplied at a faster rate
than in the gas models to
stop and reverse the core collapse and then establish post-collapse expansion.
Subsequently, as with the rate of binary formation,
the rate of energy generation during the advanced post-collapse phase is
virtually the same for both $N$-body and gaseous models. What has just
been said about models with continuous energy
generation is also true for those with stochastic energy generation, but during
the collapse
phase the stochastic models give much better agreement with the $N$-body data.

\medskip
{\bf 2.4  Interactions of Binaries}
\smallskip

Since our data is recorded only once each time interval, and since we
have little information on the identity of binaries present at each time,
it is difficult to present unambiguous data on  such problems as the
changes in energy experienced by binaries in interactions.  However,
since it is unlikely that there is more than one very energetic binary
present in each model at each time (\S2.3), the results
should give a good guide
to {\sl its} behaviour.  For example, a study of the energy changes (between
successive time units) above a suitable threshold
indicates an average relative change of order
$0.2$ or less, though with much scatter, especially for the most energetic
pairs. The value $0.2$ is predicted by the binary scattering cross section
given by Spitzer (1987), which is itself rather close to that found numerically
by Heggie \& Hut (1993).  In our view this value of the mean relative energy
change supersedes the value of 0.4 which was originally suggested by
scattering experiments at low impact parameter (Hills 1975) and the theory
of close interactions (Heggie 1975).  The mean relative change in
binding energy per interaction is relevant to the
energy which a binary reaches before being liable to escape and to
the total energy which each hard binary yields before escaping (\S6.3).

The scattering cross sections just mentioned predict a steep fall-off
with increasing $\Delta$ (the relative increase in binding energy), and
considerably steeper than found from our data.  This may occur because
more than one weak encounter may occur within our sampling interval,
which suppresses the correct abundance of encounters with low values of
$\Delta$. Better statistics on these questions, again made by direct
observation of $N$-body systems, can be found in McMillan (1989).

The larger energy changes are responsible for ejecting binaries to large
radii (Hut \et 1992).  Our results show that
there is an increasing {\sl spread} in the spatial distribution of hard
binaries with time.  While there is almost always at least one binary
within the core in the post-collapse phase, the mean distance of the
most outlying one increases roughly comparably with the half-mass radius.

Though the binaries may be found over a wide range of radii, their
interactions take place when they are in the regions of higher
density.  Almost all interactions (defined as any recorded change in
the binding energy above an arbitrarily chosen threshold, when there
was no ambiguity in the position of the binary) take place within the
half-mass radius, and at least half within the core radius.

The above discussions on the spatial distribution of hard binaries
within the cluster and the locations of their interactions raise the
following question: Where does a binary deposit the energy it emits in
an interaction?
The answer to this question depends on the relation between the escape
velocity from the cluster and the escape velocity from the core. The
smaller the system the smaller is this ratio (see Fig.20, which
portrays the central potential scaled by the central velocity
dispersion), and the easier it is to remove stars and binaries from
the core, and to disperse them widely throughout the cluster. This
means that for low $N$ the energy generated by binaries is more widely
distributed within the cluster than for high $N$, where it should be
relatively more concentrated within the core. This difference is
reflected in the spatial evolution of the cluster. The intermediate
and outer Lagrangian radii evolve considerably faster for $N = 250$
than for larger models even after allowance is made for the difference
in relaxation times (see Giersz \& Spurzem 1993 for more discussion).
The slopes of the profiles of density and velocity dispersion are also
markedly different for models with low and high $N$ (cf. \S\S3.3,
4.2).

The significance of this discussion is that any formula of the type presented
in eq.(12) cannot correctly represent the
spatial and temporal distribution of energy generation in $N$-body models,
though results in later sections of this paper suggest that it can be treated
as a reasonable compromise between the real complexity of binary interactions
in $N$-body systems and the need for a simple
formula which can be used in gas or Fokker-Plank models.
\bigskip

{\bf 3.  SPATIAL EVOLUTION}
\medskip
{\bf 3.1  The Half-Mass Radius}
\smallskip

As mentioned in \S2.2, orthodox theory implies that the post-collapse
evolution of the total energy of a system is determined by average
properties of the whole cluster, and not by dynamical processes in the
core. We follow this distinction by discussing separately the evolution of
the inner Lagrangian radii (those which contract during core collapse)
and the outer parts of a cluster.  For the latter we concentrate on the
half-mass radius.

Studying the $N$-dependence of this evolution is complicated by the fact
that our models reach rather different stages of post-collapse
evolution, in a rather $N$-dependent way.  Our clusters for $N=250$ are
virtually complete for about $11$ collapse times, whereas for $N=1000$
they stop at about $2$ collapse times.

For $N = 250$ the late evolution is sufficiently far advanced that
loss of mass by escape (cf. \S6) has had a significant effect.  Let us
parameterise this by supposing that the bound mass $M$ varies with $t$
as $$ M(t)=a(t-t_0)^{-\nu}, \eqno(14) $$ (Goodman 1984), where $t_0$
and $\nu$ are constants.  Assuming that $t-t_0$ is proportional to the
half-mass relaxation time, and neglecting the variation with time of
the Coulomb logarithm, these simple considerations lead to the result
$$ r_h(t)=b(t-t_0)^{(2+\nu)/3}, \eqno(15) $$ where $b$ is another
constant.

Empirically, by fitting the dependence of the bound mass and half-mass radius
on $t$, we can find values of $\nu$. Unfortunately these values depend on
the range of $t$ over which the fit is carried out. We found that by
combining eqs.(14) and (15) we can get more consistent results by fitting
the half-mass radius as a function of bound mass, i.e. the relation
obtained by eliminating $t$ in the above equations. The values of $\nu$ for all
models are presented in Table 3.

The theoretical form of eqs.(14) and (15) corresponds to the
self-similar gas model
computed by Goodman (1984) and, if we ignore $\nu$, to the Fokker-Planck
model of H\'enon (1965).  Comparison of the parameter $a$ or $b$ in
eqs.(14) and (15) with
the corresponding forms in these models permits a determination of the
parameters $\gamma$ in the Coulomb logarithm and also a constant $C$ which
appears in the gas model in the coefficient of thermal conductivity
(Lynden-Bell \& Eggleton 1980). Note that we have chosen a different
method of determining these parameters from the method used in the
analysis of core collapse in Paper I.  There are reasons for supposing
that the best values of the parameters $\gamma$ and $C$ will be
different in collapse and post-collapse, and so the latter values cannot
be determined by an overall scaling of the results of an evolutionary
gas- or Fokker-Planck model (i.e. a single scaling valid for both pre-
and post-collapse evolution), which would be the closest analogue to the
method adopted in Paper I. The best values we have deduced are given in
Table 3 with subscript $1$, but note that  $\gamma$ was determined
solely by
comparison with H\'enon's model (taking $\nu = 0$), and then $C$ was
determined from Goodman's model using this value of $\gamma$. Assuming that
H\'enon's model could also have  the same $\nu$ dependence as Goodman's model
if allowance were made for mass loss (i.e. a power-law index
equal to $(2+\nu)/3$ for the dependence of the half-mass radius on time,
instead of $2/3$)
we can estimate new values of $\gamma$ and $C$, which are given in Table 3
with subscript 2.

Perhaps, however, the best way of estimating these parameters is to
eliminate $\nu$ from eqs.(14) and (15) and fit
the resulting formula
$$
M^{1/2}r_h^{3/2}=c(t-t_0), \eqno(16)
$$
to the $N$-body data. The coefficient $c$ is a known function of
$\gamma$ for H\'enon's model, and of $\gamma$ and $C$ for Goodman's
model, and so we can determine them using values of $c$ derived for each $N$.
The best overall values deduced in this way are $C = 0.164$ and
$\gamma = 0.035$.

The value of $C$ can also be estimated, without any reference to
$N$-body models whatever, by direct comparison between H\'enon's
and Goodman's models, provided that $\nu$ can be set equal to zero
in the latter. The result depends on the value of the scaled half-mass
radius in Goodman's model,
which is a weak function of $N$. For the limit of low $N$ we
found that $C$ is equal $0.158$.

The consistency of these values gives an indication of how well the
$N$-dependence of the evolution is captured by these models.  Note
that the value of $C$ is about 50\% larger than the best value found
for core collapse (Paper I). However the latter value was determined
mainly by analysis of the inner parts of the systems. As was mentioned
in Paper I, comparison between gas and Fokker-Planck models indicates
that the effective value of $C$ is larger for the outer parts of a
system during core collapse. Also, a determination of $\gamma$ and $C$
for the half-mass radius of $N$-body models during core collapse
suggests values of $0.05$ and $0.125$, respectively. Therefore the
values found for the post-collapse evolution do not differ so greatly
from those obtained for the same Lagrangian radius in the collapse
phase.  Unfortunately, determination of the best values of $C$ and
$\gamma$ for the inner Lagrangian radii in post collapse evolution
could not be carried out in the same way as for the half-mass radius,
because we do not know the values of the scaled inner Lagrangian radii
for Goodman's models (Goodman 1984).  Nevertheless we believe that the
values of $C$ and $\gamma$ do change in the post-collapse phase. $C$
should be bigger and $\gamma$ smaller than in the collapse phase, but
the appropriate values may depend also on which Lagrangian radius is
being considered.

Closely associated with the evolution of the half-mass radius is the
total energy of the bound members of the system, if we exclude the
internal binding energy of bound binaries.  This defines what Aarseth
\& Heggie (1992) termed the ``external" energy of the system, and it is plotted
in Fig.3.  Here the models for $N = 250$, $500$, $1000$ and $2000$ are plotted
together, using the $N$-dependent scaling deduced in Paper I; though
this scaling is based on the behaviour of the collapse phase, whereas we
have found a different value for $C$ after core collapse, this does not
affect the relative location of the three curves on this plot.  What is
evident here is, once again, the later production of energy in core
collapse for larger $N$ (cf. Fig.2), whereas the production of energy
becomes more and more closely synchronised after core collapse, in
agreement with H\'enon's argument.
\medskip

{\bf 3.2  The Inner Lagrangian Radii}
\smallskip

Before we discuss the interpretation of the spatial evolution of the
inner parts of the cluster, we present a typical result (Fig.4) which
summarises the entire evolution.  It is clear visually that the
behaviour of the systems is close to being self-similar throughout
almost all of the post-collapse expansion, at least as far as the half-mass
radius. (The radii containing the innermost $75\%$ and $90\%$ of the mass
expand slightly faster, however.)  This nearly self-similar evolution is
exhibited by our models with $N = 250$, $1000$ and $2000$ as well.
Another way of presenting this data, though it is not shown here, is to
compute a logarithmic density profile (based on the mean density between each
Lagrangian sphere) plotted against a radial variable scaled to $r_h$:
after core collapse the profiles are virtually identical except for a
time-dependent vertical shift.

According to standard theory, in post-collapse expansion the core
adjusts itself so that energy is released there at just the rate
required to support the expansion of the outer parts of the system.
Therefore an adequate theoretical explanation of the behaviour of the
inner Lagrangian radii depends on successful modelling of the mechanism
of energy generation, which in $N$-body models is usually thought of as
being due to the formation, evolution and expulsion of binaries formed
in three-body encounters.

In the first place we compared the evolution of our $N$-body models with
that of a gas model in which binaries yield energy at a rate given by
eq.(13).  For $N = 250$ this was found to make the minimum of the
innermost Lagrangian radii (at ``core bounce") too shallow; a good fit
here required a reduction in the coefficient to about 25, for which we
could find no plausible explanation.

Note that eq.(13) assumes that binaries are formed at a certain rate,
and instantaneously emit a certain amount of energy.  Two obvious
shortcomings of these assumptions are that both the formation and the
evolution of a binary are stochastic.  In fact it is not difficult,
within the scope of a gas model, to incorporate these effects, as has
been done by Takahashi \& Inagaki (1991) for the Fokker-Planck model.
For the record we state here the formulae we used to model the
stochastic effects.  In each time step the formation rate of permanent
binaries was computed from the formula of Hut (1985), and a
pseudo-random number was used to decide whether a binary forms.  If so
it is added to a list of existing binaries.  In each time step each
binary is tested, again probabilistically, to decide if it experiences
an encounter with a single star.  For this purpose the scattering
cross sections of either Heggie (1975, eq.(5.65)) or Spitzer (1987,
eq.(6-27)) were used, the location of the binary being assumed to be
the centre of the system.  In the event of an encounter, a fraction of
the energy released is added to the core. (In gas models we cannot
follow movements of binaries and single stars in the system, and
therefore we do not know where the energy released by binaries is
deposited. To overcome this problem we introduced a parameter which
describes what fraction of the total energy generated by binaries is
deposited in the core. If it is smaller than one the remaining energy
is assumed to have no influence on the system evolution.) If the
increase in the translational speed of the binary takes it above the
escape speed from the centre (assuming that the initial translational
speed is $v_c/\sqrt{2}$, where $v_c$ is the central root-mean square
three-dimensional velocity), the binary is removed from the list.

We found that stochastic models  gave a deeper
minimum of the inner Lagrangian radii at the end of core collapse than
continuous models.  In
fact it was rather {\sl too} deep, and a possible explanation for this
lies in our technique of averaging results for a number of $N$-body and gas
systems. The spread of times of core bounce for $N$-body models is bigger than
for stochastic gas models due to  fluctuations caused by the small numbers of
stars in the cores of the $N$-body models.
Because of this spread, the average Lagrangian radii for $N$-body models will
reach  shallower minima than stochastic models.

Except for this feature, the stochastic models can be adjusted to
provide a very good fit to all properties of the
core (core radius, central density and velocity dispersion) for the whole
evolution. We found that a stochastic model in which about half of the total
energy generated by binaries is deposited in the core,
with Spitzer's scattering cross
section, gave the best fit to the N-body data. Other models (e.g. Spitzer's
scattering cross section with all the energy generated by binaries being
deposited in the core, or Heggie's scattering cross section) gave values of
core parameters which were systematically in error, especially around core
bounce. However, we would like to stress that all stochastic models which we
tried gave reasonably
good fits for the post-collapse evolution. In this phase the evolution is
monotonic, and one might expect that stochastic effects would play a less
essential role. Therefore we have also considered how well a gas model based on
eq.(13) fits the $N$-body data. For $N = 250$ the data are fitted
best if the value of
$C_b$ is reduced to about $55$,
though higher values, nearer $70$ and $90$, are preferred for $N = 500$
and $N \ge 1000$, respectively (cf. Giersz \& Spurzem 1993).

Several factors
contribute to the value of this coefficient, and it is not immediately
clear which one has to be modified to account for our empirical finding
that, for small $N$,
a smaller value is needed than the one usually adopted (i.e. $C_b = 90$,
cf. Cohn 1985).  Among these factors are the following:
(i) the heating may be less localised than is implicitly assumed
in eq.(13); (ii) the total energy emitted by a binary before it
escapes depends on the mean energy emitted in each encounter: eq.(13)
with $C_b = 90$
is based on a value $0.4$ for the mean relative change in energy (Heggie 1975
and Hills 1975),
but we have already seen (\S2.4) that there are reasons for supposing
this is too high; (iii) the energy emitted per binary is related to the
central potential, and in the systems with small $N$ this is smaller (relative
to the velocity dispersion) than is usually assumed (see \S6).  Explanations
(i) and (iii), which are related, predict a dependence on $N$
which is consistent with our $N$-body data. For more discussion on this issue
we refer to the paper of Giersz \& Spurzem (1993).

Our assertion that stochastic phenomena have little effect after core
bounce may apply to average results for an ensemble of systems,
but it does not apply to individual cases.  Fig.5 shows results for one
particularly prolonged calculation. It exhibits a small number of
episodes in which the inner Lagrangian radii show abrupt expansion (by as
much as a factor of two) followed by slow recollapse.  It is customary
to ascribe such behaviour to the loss from the core of a binary (by
recoil after an energetic interaction), but what puzzles us about
results such as this is that relatively few such interactions are
accompanied by this kind of abrupt expansion, and those that are are not
necessarily the most energetic and spectacular ones.

It is clear that these episodes of collapse and reexpansion
(superimposed on the long-term post-collapse expansion) can
occur at different times in different systems.  For this reason it turns
out that the dispersion in the Lagrangian radii (over a given ensemble
of models) is much larger in post-collapse evolution than during core
collapse. For all Lagrangian radii during the post-collapse phase the
dispersion is nearly a constant fraction of radius.
These fractions increase with decreasing radius, reaching values of about $0.5$
for the innermost Lagrangian radius (in the case $N = 500$).

\medskip
{\bf 3.3.  Logarithmic Gradient Profiles}
\smallskip

At the end of this section we would
like to present our results on the spatial evolution of $N$-body systems
in the form of profiles of the logarithmic density {\sl gradient}. One might
think that this would be of no value because of the statistical noise in
the $N$-body data.  We shall see, however, that this may be overcome by
the simple techniques of combining data from large numbers of
independent simulations, and time-smoothing.  In another sense what we
shall be doing is presenting the data of Fig.4 in a different way, and
we shall discuss corresponding results for other $N$.

Our $N$-body models provide information about the radii of Lagrangian
shells, and not information on the density directly.  In order to
convert the available data into profiles of density gradients, let us
assume that the density can be expressed by the function $\rho = A
r^{-\alpha}$ within two successive Lagrangian shells.  By using this
formula to compute the masses of the two shells (i.e.  between three
consecutive Lagrangian radii) and comparing with the $N$-body data, it
is possible to estimate the value of $\alpha$, the logarithmic density
gradient, at the middle radius.  However, in order to reduce the
statistical ``noise'' of the $N$-body data further, it was first
smoothed using the standard routine SMOOFT from Numerical Recipes
(Press \et 1986), and then the procedure described above was
implemented.

To check the validity of our method we used it to compute the
logarithmic density gradient profiles for some isotropic and anisotropic
gas models (Spurzem 1993), i.e.  profiles based on the Lagrangian radii
of these models, and compared them with the values given directly by the
models.  The results obtained by both methods agreed very well.

Turning now to the profiles obtained from our $N$-body data, we find
that the results vary rather markedly with $N$
(see Fig.6a,b). For $N = 250$ the core collapse
does not proceed to sufficient depth to develop the structural characteristics
of self-similar gravothermal collapse, i.e. a constant value of $\alpha$ for
Lagrangian radii around the collapsing
core (Cohn 1980). It seems that core collapse does not enter
this self-similar phase at all. The post-collapse profiles agree very
well, however, with a homologous
expansion scenario, as all profiles are very similar to that at the time
of core bounce. By contrast, the characteristic constant value of
$\alpha$ is clearly seen in a {\sl gas} model for $N = 250$ (Fig.6c).

For bigger systems the $N$-body models develop deeper collapse, and
the logarithmic gradient profiles of density become more and more
similar to those for gas models. The profiles for $N = 2000$, for both
$N$-body and gas models, are presented in Figs.6b and 6d,
respectively.  What is evident from Fig.6d is the decrease of $\alpha$
(increase of $d\ln\rho/d\ln r$) at intermediate Lagrangian radii, both
for very advanced core collapse and for homologous post-collapse
expansion. Similar behaviour for $N$-body data is clearly visible only
for $N = 10000$ (Spurzem 1993).  This tendency is qualitatively
consistent with the post-collapse homologous models of Inagaki \&
Lynden-Bell (1983) at least for a limited range of radii.  Real
systems with smaller $N$ do develop homologous expansion but their
spatial structure is more complicated than simple theory predicts.

The technique described above provides useful additional information on the
spatial structure of $N$-body systems, and in future it could be used to
examine more realistic $N$-body calculations of systems with larger $N$.

\bigskip

{\bf 4.  EVOLUTION OF THE VELOCITY DISTRIBUTION}
\medskip
{\bf 4.1  Comparison with Continuum Models}
\smallskip

With regard to the distribution of stellar velocities,
the data discussed in this paper have little to
add to what was said in Paper I, until late core collapse. Therefore
in this section we concentrate on
the velocity dispersion at core bounce and in the post-collapse phase.
Figs.7 (a) and (b) are relatively typical of all our results in the
inner and outer parts of the cluster, respectively.  It is especially
important in late core collapse to exclude the ``internal" kinetic
energy of
binaries, especially hard ones, and in our computations this was done as
follows. For regularised binaries only the velocities of the
centres of mass contributed to the average velocity dispersions between
successive Lagrangian radii. For non-regularised binaries, which
are typically much softer than regularised
ones, the internal (relative) motions of the components were not subtracted.
This should not have any serious effects on the determination of the velocity
dispersion because these binaries have internal speeds less than
or comparable
to the velocity dispersion itself.

First we discuss the velocity dispersion beyond the half-mass radius.
This agrees quite well with the predictions of an isotropic gas model
for the collapse phase, but during the post-collapse phase the
disagreement between the models increases with time. The discrepancy
first becomes noticeable at about the time of core bounce, and can be
associated with the development of anisotropy in the outer parts of
the system (see \S 4.3 below and Giersz \& Spurzem 1993). The
anisotropy $A$ reaches values of about $1$ at time $t\sim1000$ (for $N
= 500$) and fully accounts for the differences between the models at
this time.  The occasional `spikes' visible in Fig.7b have a different
origin: examination of individual cases shows that the velocity
dispersion changes abruptly during passage of energetic escapers.
Although the details of this process are suppressed in Fig.7(b), which
averages results over large numbers of cases, we can still see a few
spikes which are associated with very energetic products of
interactions between stars and binaries.

For the gas model shown in Fig.7(b) the results are
insensitive to the treatment of binaries, but this becomes relatively
more important in discussion of the inner shells of the models.  This is
illustrated in Fig.7(a), which shows how the velocity dispersion evolves
for stars between Lagrangian radii corresponding to the innermost $1\%$
and $2\%$ of the mass.  The isotropic gas model provides satisfactory
agreement with the $N$-body results in the
post-collapse phase, at least well after core bounce. At around the time of
core bounce gas models with stochastic
binary heating provide quite satisfactory fits to the $N$-body data, but gas
 models
with continuous energy sources give too low a velocity dispersion. This
confirms the conclusion of \S3.2
that stochastic effects are very important during core bounce.  (The gas
model whose results are plotted in Fig.7a incorporates steady heating
which is, however, only included after $t=130$, cf. Giersz \& Spurzem
(1993).  For the value of $C_b$ see \S3.2).

The temporary episodes of collapse and reexpansion of the core, referred
to in \S3.2, also contribute a
signature, though of smaller amplitude, in the central velocity
dispersion, when this is plotted for individual  cases.

\medskip
{\bf 4.2  Evolution of the Velocity Profile}
\smallskip
An alternative view of the evolution of the velocity dispersion is
exhibited in Fig.8, which shows velocity dispersion profiles at various
times.  Since our data is relatively crudely binned by radius these
profiles have relatively little detail, especially in the outer zones.
Within the inner zones, on the other hand, where the number of stars is
smaller, the results are subject to greater statistical uncertainty, and so
these plots were generated by averaging over 5  successive times.
The mean square speed within a Lagrangian shell is scaled by the central value
of the mean square velocity and plotted at
the mean radius of the shell, scaled by the half-mass radius.

Despite the crudeness of these profiles, what is clear is how closely
self-similar they appear to be in the post-collapse phase of the
evolution.  Indeed the lines for post-collapse evolution are almost
indistinguishable, except for fluctuations.

The technique of computing the logarithmic density gradient (\S3.3) can
be extended to allow the computation of the profile of the logarithmic
velocity dispersion gradient.
The results are
rather similar to the corresponding profiles of gaseous models, except
in the outer parts of the system (i.e. radii beyond the half-mass
radius). During the post-collapse evolution the $N$-body models develop
a somewhat smaller gradient of velocity dispersion than gas models do. This
may be
connected with the development of anisotropy in the $N$-body models, although
anisotropic gas models of Spurzem (1993) also failed to predict this feature.
However, as  was stressed by Giersz \& Spurzem (1993), anisotropic gas
models also failed to reproduce the correct {\sl spatial} evolution of
the $N$-body models at the outermost
Lagrangian radii. Possible explanations for these discrepancies include
the non-local nature of heat transfer in $N$-body systems, which is
connected with close two-body interactions and interactions with hard
binaries.  These affect the distributions of density and velocity
dispersion at locations far from the site of the interactions, and so
cannot be modelled properly by anisotropic gas models
with local heating (cf. Giersz \& Spurzem 1993).

\medskip
{\bf 4.3  Anisotropy }
\smallskip

As a measure of the anisotropy we have adopted the definition
$A = 2 - 2\sigma_t^2/\sigma_r^2$, where $\sigma_t$ and
$\sigma_r$ are the one-dimensional tangential and radial velocity dispersions,
respectively.  Each was computed as a mass-weighted average taken over
all stars within spherical
shells bounded by consecutive Lagrangian radii.

There is no discernible anisotropy in the innermost shells up to the
Lagrangian radius for $10\%$ of the mass. For intermediate shells the
anisotropy starts to increase nearly at the time when binaries start
to influence the core evolution. For these shells it then increases
nearly linearly with time, and eventually reaches a maximum value at
about the time when the cluster enters the phase of homologous
expansion. Finally, for the outermost shells the anisotropy increases
nearly linearly from the very beginning of the calculation, and
eventually reaches a maximum value at about the same time as for the
intermediate shells.  The maximum value of the anisotropy depends on
the position within the cluster and on the total number of stars. It
increases markedly with radius, and it seems that the maximum value is
smaller for larger systems (particularly for the outermost shells).

Our results suggest that some aspects of anisotropy are closely
connected with binary activity, especially for small $N$: interactions
of binaries with single stars, and the expulsion of stars and binaries
from the core to the outer parts of the system. This conclusion is
supported by the facts that the anisotropy starts to increase in
intermediate shells at about the same time as binaries start to
influence the core evolution, and that it levels off at about the same
time as the energy generated by binaries has become adjusted to the
demands of the overall cluster expansion. A detailed discussion of all
aspects of anisotropy, and a comparison between $N$-body data and
anisotropic gaseous models, are presented in Giersz \& Spurzem (1993).

\bigskip

{\bf 5.  CORE EVOLUTION}
\medskip
To some extent the basic data of core evolution have already been
covered in \S\S3.2 and 4, and even parts of \S2.  But it is such an
important feature of star cluster evolution that it deserves an
integrated study, in which all this data is synthesised.

\medskip
{\bf 5.1  Core Bounce}
\smallskip

Of the various quantities we studied, we have found that measurement of
the central potential provides what is apparently the sharpest estimate
of the time at which core collapse may be considered to end (Fig.9).
In fact the central
potential  can  itself be
estimated in several ways, of which we have considered two: the
potential at the location of the density centre (defined in Casertano \& Hut
1985), and
the minimum (over all stars in the cluster) of the potential at the
location of each star; though in this case the contribution of the
nearest neighbour of each star is omitted, in order to exclude the
contribution from a binary companion.  On average the former measure
gives slightly lower values for the central potential, and it is slightly
noisier than the latter.

These results show rather clearly that core collapse ends relatively
later (i.e.  when the collapse is more advanced) in the larger models.
Thus the time of core bounce, $t_{coll}$, is not simply proportional to
$N/\ln(\gamma N)$, as would be the case if it were simply proportional to the
half-mass relaxation time.  In fact we have found that
$t_{coll}/t_{rh}$ is approximately $13.5,~14.1,~16.8$ and $17.3$ for $N =
250, ~500, ~1000$ and $2000$, respectively, where $t_{rh}$ is the initial
half-mass relaxation time (Spitzer 1987, p.40, except that the value of
$\gamma$ has been taken as $0.11$, from Paper I, instead of Spitzer's value
of $0.4$). The depth of the potential
well at core bounce is also greater for larger $N$.

Incidentally the data for $N=2000$ show two
minima of rather similar depth at times $t\simeq 330$ and $360$ in
Fig.9, but our other measure of central potential (see above) clearly
resolves the issue and implies that the later is the true minimum.
Another detail worth recording is the broad range of times of core
bounce exhibited by different cases with the same $N$.  Actually this is quite
difficult to determine for $N \leq 500$, but for $N \geq 1000$ the range
of values has a length about 50\% of the mean value.  To the extent that
the value can be found for smaller $N$ the absolute range is somewhat smaller,
but comparable with the mean time of core bounce. It seems that the range
in the time of core bounce, expressed as a fraction of the mean time of core
bounce,
decreases with increasing $N$, as is certainly plausible on the simplest
grounds. Unfortunately, however, we cannot quantify this statement
because of substantial uncertainties in the determination of the time of core
bounce for systems with $N \leq 500$.

Theoretically, conditions at core bounce are usually estimated by
assuming that the energy emitted by binaries on the time scale of the
collapse is comparable with the energy of the core (Heggie 1984,
Goodman 1987).  This leads to the condition that $$
N_{cb}\propto\left({ f
\vert\phi_c\vert\over\sigma_c^2\xi\ln\Lambda}\right)^{1/2}\eqno(17) $$
(cf. eq.(12)), where $\xi$ is the dimensionless core collapse rate
(Cohn 1980), $f$ is a coefficient of order $1$ which determines the
efficiency of the energy sources, $N_{cb}$ is the number of stars in
the core at core bounce, $\ln\Lambda$ is the Coulomb logarithm, and it
is assumed that the main energy generation mechanism is associated
with the formation and evolution of binaries in three-body encounters.
Assuming that the core density and radius ($\rho_c$ and $r_c$) are
related by $\rho_c\propto r_c^{-\alpha}$, as in the standard theory of
core collapse (Lynden-Bell \& Eggleton 1980), this leads to the
condition (at core bounce) that $$
\rho_c\propto \left({\xi N^2\ln\Lambda \over f\vert\phi_c\vert/\sigma_c^2}
\right)^{\displaystyle{\alpha\over2(3-\alpha)}}, \eqno(18)
$$ where as usual we use standard units (in which the total initial
mass of a system is unity).  The results of Fig.10 and Table 4,
however, indicate that the central density at the time of core bounce
is not a simple power of $N$ within the range we have studied.  Within
the context of the above theory, this indicates that one or more of
the parameters in eq.(18)
are variable.  Indeed Fokker-Planck models indicate that $\alpha$ and
$\xi$ do evolve as the core collapses (Cohn 1980), and we have already
seen (Fig.9) that the value of $\phi_c$ at the time of core bounce
depends
on $N$.  Indeed the character of late core collapse already looks
different for $N = 1000$ and $2000$ in Fig.10 than in the results of
the other two sets of models.

The theoretical interpretation of the density at core bounce is actually
somewhat more involved than the above discussion suggests.
One complication, already mentioned in \S3.2, is the effect of averaging over
many cases in which the core bounce can occur at very different times.  It
can be easily appreciated that this tends to suppress the mean density
at its peak (at core bounce). This variation of bounce
times is also exhibited in gas models with stochastic binary formation
and evolution, but not to the same extent.

What has been said above about the central density at core bounce
applies also, with appropriate changes, to the central velocity
dispersion. The $N$-dependence of
the maximum central root mean square velocity is much milder, the value
for $N=2000$ being only about $18\%$ larger than for $N=250$ (Fig.11).
Again the most successful model for binary heating is one including both
stochastic formation and stochastic burning.

We have also determined the core radii in our models according to the
following prescription: Lagrangian radii for the one or two innermost shells
(for $N \leq 500$ we used two shells and for $N > 500$ only one), together with
the average velocity dispersion in these shells, were used to estimate
the central
density and velocity dispersion. The core radius was then calculated using the
standard definition $r_c^2=9\sigma_c^2/4\pi G\rho_c$.
Because of the relatively modest time-dependence of the velocity
dispersion, the behaviour of the core radius is almost entirely
determined by that of the central density.  So, again, the evolution
is best described by a gas model including stochastic formation and
stochastic  burning of binaries. The main core parameters at the time of core
bounce are presented in Table 4.

A relatively direct test of the theoretical argument leading to eq.(17)
is in the sixth column of Table 4.  It shows that the number of stars in
the core at core bounce (i.e. the number within radius $r_c$)
is not independent of $N$, as often assumed, but
increases with increasing $N$, at least within the range we have
studied. The trend is consistent with eq.(17) in the sense that, for
larger $N$, collapse would be expected to proceed further (e.g. in terms
of central density); and it is known that $\xi$ decreases and $\vert \phi_c
\vert/\sigma_c^2$ increases
as core collapse deepens (Cohn 1980).

\medskip
{\bf 5.2  Post-Collapse Evolution of the Core}
\smallskip

For the purpose of theoretical comparison the simplest choice is the
series of gas models computed by Goodman (1987), which assume a simple model
of energy generation by binaries, and self-similar evolution.  Goodman's
models are stable for the range of values of $N$
considered here.  The inner Lagrangian radii in our $N$-body results, at
least
up to the $75 \%$ radius (Fig.4),
approximately confirm the homologous nature of the evolution.  Other gas
models, without the assumption of self-similarity, have also been
computed in order to determine the effects of different assumptions on
the rate and character of energy generation by binaries.

The actual value of the central density is quite sensitive to the
assumed rate of energy generation by binaries, and what was said about
eq.(13) in the context of the inner Lagrangian radii in \S3.2 is
applicable here;  in other words this formula, with a suitably adjusted
numerical coefficient, provides a good fit in the post-collapse regime,
though inclusion of stochastic effects is needed for a satisfactory fit
at core bounce.   Similar remarks may be made about the central velocity
dispersion.
Indeed, when the time units are adequately
scaled (Fig.11), the differences (for different $N$)
which appear at around the time of core
bounce rapidly diminish.

Since the core radius depends solely on the central density and
velocity dispersion, we also find that its post-collapse evolution
(substantially after core bounce) fails to discriminate the models
with stochastic binary formation and burning from those with smooth
burning.  There is, however, one interesting feature which may
indicate the role of stochastic binary evolution, which is a slight
dip in the otherwise monotonic increase in the core radius.  For $N =
250$ it occurs at about $t\simeq 160$; it may also be glimpsed in
Fig.10 as a slight rise at the scaled time of $t\sim 500$ for $N=250$.
It is possible that this marks a tendency for recollapse of the core
at a time when the first energetic binary is ejected from the core.
This time will vary greatly from one case to another, but the signal
of this process may appear weakly when many cases are averaged.  A
similar feature may be present in our results of models incorporating
stochastic burning and formation of binaries, but in any case is
somewhat elusive.  What is not in question is the evidence for this
recollapse in individual cases (e.g. Fig.5).

In the homologous phase after core collapse, the
$N$-dependence of the core radius qualitatively
follows theoretical predictions,
in the sense that, for larger $N$, it is a smaller fraction of the
half-mass radius (see Fig.12).
The number of stars in the core in the early post-collapse phase tends
to stay fairly close to the value at the end of core collapse (Fig.13), and so
the $N$-dependence of $N_c$ at that time approximately persists for a
time comparable with the collapse time itself; and indeed for a further
11 and 6 collapse times for $N = 250$ and $500$, respectively.
We did not follow the
evolution of the cases with $N = 1000$ for so long, relative
to the collapse time, and so the long-term evolution of $N_c$ is unclear
in this case.
{}From Figs.12 and 13, however, it seems that
core bounce has a visible influence for larger $N$: the ratio of $r_c/r_h$ and
the number of stars in the core at the time of core bounce are slightly smaller
then during the post-collapse phase, for $N = 2000$. This is in agreement with
theoretical predictions, which, in the limit of large $N$, predict that
the value of $N_c$ at core bounce should be almost independent of $N$
(eq.17), whereas in homologous post-collapse expansion it varies as
$N^{1/3}$ (eq.8). For models with lower $N$ this feature is not present,
because the core collapse does not proceed far enough and, as we saw in
\S5.1, the $N$-dependence of eq.(17) is complicated by other factors.
The number of stars
in the core predicted by Goodman's gas model (Goodman 1987) is rather
too large for $N \leq 1000$, but for $N = 2000$ it is very close to that found
in the $N$-body calculations.

\medskip
{\bf 5.3  The Evolutionary Track of the Core}
\smallskip

The foregoing discussions have considered the time-dependence of various
quantities, but it is also of interest to consider the variation of two
(or more) fundamental core parameters against each other.  For example, in the
context of gas and Fokker-Planck models this has proved to be a useful
diagnostic for the study
of oscillatory post-collapse evolution (Goodman 1987, Breeden \et 1990).

Fig.14 shows results for $N = 500$ and a gas model with continuous
energy generation at a rate which has been found empirically to give
satisfactory agreement with core parameters in post-collapse evolution
(see \S3.2).  As we have already observed, this model does not produce a
sufficiently deep collapse, but the post-collapse evolution of the core
is followed quite closely.

The post-collapse behaviour shown in Fig.14 is reasonably well explained by
the usual theory of homologous post-collapse evolution. If we suppose that
the velocity dispersion varies as $v_c^2\propto GM/R$, where $M$ is the
total mass and $R$ is a scale radius, and that the core density varies
as $\rho_c\propto M/R^3$, then we find that
$$
{d\ln v_c^2\over{d\ln \rho_c}} \simeq {1\over3}
\left(1-{2\over3}{d\ln M\over d\ln R}\right).  \eqno(19)
$$
Since we find (from eqs.(14) and (15) and Table 3)
that $d\ln M/d\ln R\simeq -0.128$ for $N = 500$ it follows that
the expected slope in the $N$-body data should be about $8\%$  steeper than in
the gas model (in which, by assumption, no mass is lost). This can be
confirmed qualitatively  in Fig.14.

\bigskip

{\bf 6. ESCAPE}
\medskip
{\bf 6.1  Rate of Escape}
\smallskip
In homologous post-collapse evolution it is easy to show that the rate
of escape in two-body encounters should give rise to a power-law dependence
of the total mass on time, as in eq.(14) for some value of $\nu$.
Note that this mass loss is not included in the detailed theoretical
post-collapse models which we have used up to now, i.e. those of H\'enon
(1965) and Goodman (1987).  Goodman (1984) does indeed discuss and include
mass-loss, but it is the mass loss which is associated with the
mechanism for powering the post-collapse expansion.

We find from the results of our $N$-body models that eq.(14) fits all
models with $250\le N\le 2000$ with values of $\nu$ given in
Table 3. A suitable average value for the power index is $\nu = 0.086$.
These results are
suitable for theoretical purposes, but do not give a good intuitive feel
for how quickly clusters lose mass.  Fig.15 therefore gives the raw
data.

{}From what was said in Paper I about the changing rate of escape during core
collapse, two factors contribute to the success of eq.(14):  one is that the
core evolves nearly homologously with the half-mass radius, and the
other is the near-constancy of the anisotropy in the post-collapse phase
(as far as we have followed this). As in Paper I we have attempted to
reproduce the observed rate of escape by computing the rate of escape
in an isotropic Fokker-Planck model using H\'enon's general formula
(eq.(2) in Paper I), and correcting for anisotropy in the very
approximate way discussed in Paper I (\S 3.3.3).  This is fairly successful
for all $N$, but the goodness of fit depends on the assumed escape radius
for $N$-body models (see discussion in Paper I).
\medskip
{\bf 6.2  Energy of Escapers}
\smallskip
Though the theory just described is partially successful in accounting
for the number of escapers, it grossly underestimates the energy they
carry off in the post-collapse regime.  (Here we refer to the
translational kinetic energy of single stars and the barycentres of
escaping binaries and other bound subsystems.)  The actual data are
given in Fig.16, and immediately after the end of core collapse the
values increase much more sharply than predicted by a theory based on
escape by two-body interactions.

Fig.17 helps to explain in what way the theory of two-body escape, which
accounts quite well for the fluxes of both mass and energy during core
collapse, fails in the post-collapse phase.  From the end of core
collapse onwards, a new class of escapers of much higher energy occurs.
Their numbers are relatively small, but their energy is higher by about
two orders of magnitude, and they dominate the energy flux.  It is
natural to associate them with three-body interactions, involving the
hardening of binaries.  A histogram of the energies of all escapers
(from time $t = 0$ until the end of the calculations)
shows a bimodal form, but it is clear that the distribution is
time-dependent.

To clarify this point we have carried out a Monte-Carlo
simulation of interactions between binaries and single stars using Spitzer's
and Heggie's scattering cross sections (see \S 2.4).   Briefly, a binary
is created with an initial binding energy $\varepsilon \ge 3kT$ distributed
according to $f(\varepsilon) \propto \varepsilon ^{-9/2}$ (Heggie 1975), and
allowed to evolve according to the
appropriate cross section.  Assuming that the central potential (scaled
by the
central one-dimensional velocity dispersion) is constant (see Fig.20)
the escape energies of all escapers are recorded, until the
binary itself escapes.  The number of trials computed was of order $10000$,
which is about $20$ times larger than the total number of hard binaries in the
$N$-body systems.
We have found that the maximum escape energy is as large as $1500kT$,
while the minimum escape energy is as small as $10^{-4}kT$.  Both
extreme values depend only slightly on the assumed scattering cross
section and scaled central potential, and they are in satisfactory agreement
with our experimental data, when allowance is made for the number of trials.

The presence of these escaping particles gives rise to a small virial
imbalance.  For $N = 250$, for example, the virial ratio (defined as
the ratio of the external kinetic to external potential energy) remains
within the range $0.50-0.51$ during core collapse, but then rises
abruptly over the next collapse time.  Thereafter the average value
remains close to about $0.55$, though with increased fluctuations.

\medskip
{\bf 6.3  Binary Escapers}
\smallskip

Associated with the escape of the most energetic single escapers are
escaping binaries (Fig.18).  This diagram illustrates particularly well
the value of averaging results from many simulations: such small numbers
are involved that the results from an individual case are dominated by
the stochastic nature of binary evolution.  The numbers of these
escapers are negligible compared with the total numbers of escaping
stars (Fig.15), which are therefore completely dominated by single
escapers.

Using this data, along with information on the evolution of the
half-mass radius or the external energy of the bound cluster, we can
test one of the basic assumptions of simple theory, in
which the escape of each binary is assumed to be accompanied by the
release of an amount of energy which is a given multiple of the mean
individual kinetic energy of the bound stars (Cohn 1985, Goodman 1984).
This assumption implies
that $d\ln \vert E_{ext} \vert /dN_{besc}$ is constant for each $N$, where
$N_{besc}$
is the number of binary escapers and $E_{ext}$ is the external energy of the
bound cluster. We found values: $-0.69$, $-0.46$, $-0.32$ and $-0.21$ for
$N = 250$, $500$, $1000$ and $2000$, respectively. If, as is often assumed,
each binary releases a fixed multiple of the mean kinetic energy of a single
star, the result would be proportional to $N^{-1}$. In fact the $N$-dependence
is complicated by the following fact. For larger $N$, core collapse proceeds
further, and the value of the central potential (scaled by either the global
or central value
of $kT$) is larger (during homologous expansion) than for smaller $N$
(cf. Fig.20). Hence, on average,  binaries reach larger energies
(scaled by $kT$) before escaping
(cf. Fig.2).

Though the {\sl number} of escaping binaries is relatively insignificant, their
external energy (i.e.  the translational
kinetic energy associated with their barycentric motion) is a larger
fraction of the energy of all escapers, but even so it is smaller than the
latter by a factor of about $3.5$.  Where the escaping binaries are
energetically very important is in their {\sl internal} energy (Fig.19),
which becomes comparable with the internal energy of {\sl bound} pairs quite
early in the post-collapse phase.

An especially interesting parameter, related to the internal and
external energy and number of escaping binaries, is the {\sl mean}
change of the internal binding energy of a binary in the encounter
which leads to its escape, as this is theoretically related to the
amount of energy which each binary may donate to the cluster.

The theoretical argument goes as follows.  Assume that a binary of
internal binding energy $\varepsilon$ experiences a three-body
encounter in which its binding energy increases by $\delta\varepsilon$
to $\varepsilon^\prime$.  If the change in energy much exceeds the rms
speeds of the stars, then the energy given to the barycentric motion
of the binary is approximately $\delta\varepsilon/3$.  Therefore the
binary escapes if $\delta\varepsilon/3>m v_{esc}^2 - m v_c^2/2$, where
$m$ is the mass of a single star or binary component, $v_{esc}$ is the
escape speed and $v_c$ is the central velocity dispersion; (we assume
equipartition in the core).  By the same argument we deduce that the
relative change of energy of a binary, in an encounter leading to its
escape, satisfies $$
\Delta = {\delta \varepsilon \over {\varepsilon}} ={{ 3\left(x_{ex} +
2 {{\vert \phi_c \vert} \over{ \sigma_c^2}} -1.5 \right)} \over {
x_{in} - 3\left(x_{ex} + 2 {{\vert \phi_c \vert} \over{ \sigma_c^2}}
-1.5 \right)}},
\eqno(20)
$$ where $x_{ex}$ and $x_{in}$ are the external and internal energy of
the escaping binary, respectively, both expressed in units of
$1kT_c~(\equiv m\langle v_c^2\rangle/3\equiv m\sigma_c^2)$.  For the
mean value of the ratio $\Delta$ the value of $0.4$ is frequently
quoted, based on certain analytical estimates (Heggie 1975) which are
restricted to close encounters, or numerical estimates (Hills 1975)
which are restricted to zero impact parameter. In fact inclusion of
wide encounters reduces the average, and indeed strictly it vanishes,
because of the arbitrarily large number of distant encounters which
have a negligible effect on $\varepsilon$.  Nevertheless in eq.(20) we
are concerned only with encounters which lead to the escape of
binaries, and so we can expect that distant encounters contribute only
slightly.

In order to check this formula we require some further information,
especially concerning the escape speed.  For this purpose, we show in
Fig.20 the central potential in units of the central one-dimensional
velocity dispersion.   There is remarkable consistency during the core
collapse phase, but in this paper our emphasis is on post-collapse
evolution, where there is clear evidence of an $N$-dependence, with
larger values for larger $N$, though the dependence is no faster than
about $N^{1/6}$.  This dependence presumably reflects the fact that the
ratio of core to half-mass radius decreases as $N$ increases (cf. \S5.2).

Because of the size of our sampling interval, only for a fraction of all
binary escapers were we able to trace back to determine the
value of $\Delta$ {\sl directly}, i.e. by computing $\Delta$ from the
definition  $\Delta = (\varepsilon^\prime - \varepsilon)/\varepsilon$.
For all other cases eq.(20) was used to {\sl estimate} the value of $\Delta$.

Examination of our data shows that the relative change of the binary internal
energy, in those encounters which lead to
the escape of the binary, has a very asymmetric distribution, with a very long
tail for large $\Delta$. The distribution is peaked at $\Delta$ around $0.5$,
and the average value of $\Delta$ is about $1.0$ (cf. Fig.21).  As can
be seen in Fig.21, the ``direct'' and estimated distributions are very
similar, and so it appears that eq.(20) gives quite a good
approximation for $\Delta$. However the distribution of estimated
values is slightly too high for large $\Delta$.
A possible explanation for this difference could be the fact
that we approximate the central velocity dispersion at the time of the
encounter which leads to the escape of the binary by the value at the time when
the binary is removed from the system.  This
velocity should be generally smaller than at the time of the encounter,
since the escape often leads to a modest expansion of the core, and so
our estimates are shifted towards larger
values. Other possible explanations are mentioned below, along with a
discussion of the internal energies of the escaping binaries.

As we mentioned in \S 6.2 we conducted Monte-Carlo simulations for the
evolution and escape of binaries, using two different theoretical
scattering cross sections (see \S 2.4). For both cross sections the
calculated distributions of $\Delta$ (for the final encounters leading
to escape) have a very long tail similar to those of the $N$-body data
in Fig.21. The distribution of $\Delta$ obtained for Spitzer's formula
drops more steeply than that for Heggie's formula for $\Delta \geq 1$,
and is peaked at around $0.5$, compared with a value of $0.7$ for
Heggie's cross section. Both formulae give negligible numbers of
interactions with $\Delta \leq 0.2$ which is in contradiction with our
experimental findings. The average values of $\Delta$ for Spitzer's
and Heggie's formulae are about $1.09$ and $1.33$, respectively. We
can conclude that Spitzer's formula gives closer results to those
obtained from our full $N$-body simulations. Unfortunately we cannot
quantify this statement satisfactorily, because of the limited
statistics of binary escapers from $N$-body models.

The internal energies of escaping binaries, when expressed in terms of
the central velocity dispersion, have a distribution with a very long
tail up to around $2000 kT$, and about $9\%$ of all escaping binaries
have energies $\leq 100 kT$. The distributions are peaked at around
$125$, $175$ and $225~kT$ for $N = 250$, $500$ and $1000$,
respectively. The theoretical distributions again fail to predict the
number of binary escapers with internal energies $ \leq 100 kT$, as
they give values which are several times too small. However the shapes
of all distributions, both experimental and theoretical, are very
similar for internal energies $\geq 200 kT$, and the theoretical
distributions are peaked at nearly the same values as the experimental
ones. The distributions for Spitzer's and Heggie's formulae are
practically the same.

There are at least two possible explanations for the differences between
the theoretical and experimental
distributions of
the internal energy of
binary escapers for $\varepsilon^{\prime} \leq 100 kT$.  (These may also
contribute to the differences, mentioned above, in the theoretical and
experimental distributions of
$\Delta$ for $\Delta \leq 0.2$.)  One possible explanation is connected with
the fact that a few {\sl hierarchical binaries} escape from
the systems, i.e. binaries accompanied by a distant but bound third
body. In these cases our data may refer to the ``outer'' binary.
Another
possible explanation involves the escape of relatively soft binaries
on very elongated orbits by ordinary two body processes (Stod\'o\l kiewicz
1985). It is possible that the left hand parts of the experimental
distributions in Fig.21
are contaminated by one or both of these processes.

Despite the large fluctuations in the experimental distributions for
$\varepsilon^{\prime}$ and $\Delta$ they resemble the theoretical distributions
obtained from Spitzer's scattering cross section more closely than those
from Heggie's cross section.
Another indirect piece of evidence supporting this conclusion can be obtained
from a study of the
total internal energy of {\sl bound} hard binaries. If we assume that this
energy is mainly contributed by one very hard binary (cf. \S 2.3), it may
be
estimated from our data that its energy exceeds $250 kT_c$ on average
for all $N$, in the post-collapse phase. The theoretical
estimates give values of about $220 kT_c$ and $120 kT_c$ for Spitzer's and
Heggie's scattering cross sections, respectively. Thus it seems again that
Spitzer's formula gives better agreement with the experimental data than
Heggie's one.

\bigskip

{\bf 7. DISCUSSION AND CONCLUSIONS}
\medskip
This paper has been concerned with the evolution of isolated systems of
bodies of equal mass, in those phases of the evolution in which hard
binaries play an important role.  It is unfortunate, then, that one
quantity on which our data is vague is the {\sl number} of such
binaries (\S2.2), as we instead have data on the number of pairs which are
``regularised" in the $N$-body simulations.  Though the difference
between these two numbers may correspond to pairs which are on the
threshold between hard and soft, and may be dynamically unimportant,
this flaw in our methodology should be remedied.

The other aspect of our data, as presented here, which is extremely
limited is the dispersion of the results, i.e. the size of the
differences between different cases at the same time.  This is of
special interest in the context of the post-collapse expansion of the
core (Fig.5), which in individual cases shows alternating phases of
expansion and recollapse (\S3.2).  The relation between these types of
behaviour and the evolution of binaries is far from clear.  The naive
expectation that escape of binaries causes core expansion is too
simplistic, because there are energetic interactions which do not
appear to initiate expansions, and some expansions begin with
interactions which do not seem particularly energetic.

These issues apart, our technique of combining data from relatively
large numbers of $N$-body simulations has allowed us to investigate a
number of fundamental dynamical processes with data of relatively high
statistical quality.  Concerning interactions of hard binaries, several
lines of evidence support the applicability of the simple scattering
cross section presented by Spitzer (1987).  These include (i) study of
the changes in the internal energy of hard binaries (\S2.4), (ii) the evolution
of the core (\S3.2); and (iii) the internal energies of escaping
binaries (\S6.3).  The effect of these binaries on the core, however,
requires some attention to the following points (\S\S3.2, 5.1):  (i)
models in which energy is generated smoothly, i.e. at a given average rate,
do not account well for the depth of core collapse, and the role of
fluctuations is also indicated by a study of the formation rate of hard
binaries during core collapse (\S2.1); (ii) even in
the post-collapse phase the generation of energy by these processes
seems less efficient than is usually assumed, at least for smaller $N$ (\S3.2).

In the post-collapse phase the evolution is remarkably self-similar,
whether one looks at the density gradient (\S3.3) or the profile of the
dispersion of velocities or anisotropy (\S\S4.2, 4.3). The number of
stars in the core remains nearly constant, though it depends on $N$
(\S\S5.1, 5.2).
Comparison of the
evolution of the half-mass radius with simplified models allow a
redetermination of the values of the thermal conductivity coefficient
(for gas models) and the argument of the Coulomb logarithm in the
expression for the relaxation time.  The former, expressed in terms of
the value of $C$ (Lynden-Bell \& Eggleton 1980), gives a
best-fitting value close to 0.164 (\S3.1), which is about 50\% larger
than the value determined in Paper I from the evolution of the inner
radii during core collapse.  Similarly, if the argument of the Coulomb
logarithm is expressed as $\gamma N$, the best value of $\gamma$ is
about 0.035, less than half the value found in Paper I.

The power-law expansion of the half-mass radius is matched by the
time-dependence of the total bound mass; this varies nearly as
$(t-t_0)^{-\nu}$, where a fairly consistent value $\nu\simeq 0.086$ is
indicated for all the values of $N$ we have studied (Table 3).  Most
escapers result from two-body encounters, but from about the time of
core bounce they are joined by a small number of escapers, both single
stars and binaries, which are attributable to interactions involving
binaries.  They are clearly distinguishable by their energies (\S6.2).

It can be seen that $N$-body data of suitable statistical quality can be
very useful for ``calibrating" various dynamical processes of importance
in the evolution of stellar systems, for example, binary heating.  Such
data also demonstrate the limitations of the commonly made assumption that
binaries contribute energy continuously to the core rather than
stochastically.  $N$-body models also yield data on important aspects
of stellar dynamics which are difficult to study by simplified
techniques, e.g.  stellar escape.  These simplified techniques remain,
however, the ones which are most directly applicable to systems such as
globular star clusters, because of the relatively small values of $N$ to
which useful $N$-body models are restricted.  Eventually $N$-body models
will become the method of choice for studying the dynamical evolution of
globular star clusters, but for the next decade at least reliance must
be placed on simplified models, and for the time being one of the
important tasks of $N$-body modelling is to check the assumptions and
parameters underlying these simplified methods.

\bigskip
{\bf ACKNOWLEDGMENTS}
\medskip
The research reported in this paper was supported by the UK Science and
Engineering Research Council under grant No. GR/G04820.  We are also indebted
to J.A. Blair-Fish, B.R.P.
Murdoch and Edinburgh Parallel Computing Centre for generous assistance
with the computations.  We thank S.J. Aarseth, H. Cohn and H.-M. Lee for
kindly giving us permission to use their codes.

\vfill\eject

{\bf REFERENCES}
\medskip
\leftskip=0.0 truein
\parindent=0.0 truein

Aarseth S.J., 1985, in Brackbill J.U., Cohen B.I., eds,
Multiple Time Scales. Academic Press, New York, p.377

Aarseth S.J., Heggie D.C., 1992, MNRAS, 257, 513

Breeden J.L., Packard N.H., Cohn H.N., 1990,  Preprint No. CCSR-90-2
(Center for Complex Systems Research, Dept. of Physics, Beckman
Institute, University of Illinois at Urbana-Champaign)

Casertano S., Hut P., 1985, ApJ, 298, 80

Chernoff D.F., Weinberg M.D., 1990, ApJ, 351, 121

Cohn H., 1980, ApJ, 242, 765

Cohn H., 1985, in Goodman J., Hut P., eds, Dynamics of Star
Clusters. Reidel, Dordrecht, p.161

Giersz M., Heggie D.C., 1994, MNRAS, in press (Paper I)

Giersz M., Spurzem R., 1993, submitted to MNRAS

Goodman J., 1984, ApJ, 280, 298

Goodman J., 1987, ApJ, 313, 576

Goodman J., Hut P., 1993, ApJ, 403, 271

Heggie D.C., 1975, MNRAS, 173, 729

Heggie D.C., 1984, MNRAS, 206, 179

Heggie D.C., Hut P., 1993, ApJS, 85, 347

H\'enon M., 1965, Annales d'Astro\-phys\-ique, 28, 62

Hills J.G., 1975, AJ, 80, 809

Hut P., 1985, in Goodman J., Hut P., eds, Dynamics of Star
Clusters. Reidel, Dordrecht, p.321

Hut P., McMillan S.L.W., Romani R.W., 1992, ApJ, 389, 527

Inagaki S., 1984, MNRAS, 206, 149

Inagaki S., Lynden-Bell D., 1983, MNRAS, 205, 913

Lynden-Bell D., Eggleton P.P., 1980, MNRAS, 191, 483

McMillan S.L.W., 1989, in Merritt D., ed, Dynamics of Dense Stellar
Systems.  Cambridge University Press, Cambridge, p.207

Press W.H, Flannery B.P, Teukolsky S.A, Vetterling W.T, 1986, Numerical
Recipes. Cambridge University Press, Cambridge

Spitzer L., Jr., 1987, Dynamical Evolution of Globular Clusters.
Princeton University Press, Princeton

Spitzer L., Jr., Hart M.H., 1971, ApJ, 164, 399

Spurzem R., 1993, private communication

Statler T.S., Ostriker J.P., Cohn H.N., 1987, ApJ, 316, 626

Stod\'o\l kiewicz J.S., 1985, in Goodman J., Hut P., eds, Dynamics of Star
Clusters.  Reidel, Dordrecht, p.361

Takahashi K., Inagaki S, 1991, PASJ, 43, 589

\bigskip
\vfill\eject

\parindent=0pt
\centerline{{\bf Table 1}}
\medskip

\centerline{NUMBER OF MODELS AT THE END OF CALCULATIONS}
\medskip

$$\vbox{ \settabs \+\quad 2000 \quad & \quad 56 \quad & \quad 50 \quad &
\quad 1000 \quad & \cr
\+ \hfill $N$ \hfill & \hfill $N_{ci}$ \hfill & \hfill $N_{ce}$ \hfill &
\hfill $t_{end}$ \hfill &\cr
\smallskip
\+ \hfill 250 \hfill & \hfill  56 \hfill & \hfill 50 \hfill &
\hfill 1000 \hfill &\cr
\+ \hfill 500 \hfill &  \hfill 56 \hfill & \hfill 48 \hfill &
\hfill 1000 \hfill &\cr
\+ \hfill 1000 \hfill & \hfill 50 \hfill & \hfill 41 \hfill &
\hfill 600  \hfill& \cr
\+ \hfill 2000 \hfill & \hfill 16 \hfill & \hfill 15 \hfill &
\hfill 2300 \hfill& \cr}$$
\smallskip
Note: $N_{ci}$ and $N_{ce}$ are the initial and final numbers of
computed models, and $t_{end}$ is the final time in $N$-body units.
\bigskip
\bigskip

\centerline{\bf Table 2}
\medskip
\centerline{BINARY FORMATION DURING CORE COLLAPSE}
\medskip

$$\vbox{ \settabs \+\quad $t_{coll}$ \quad & \quad 250 \quad &
\quad 500 \quad & \quad 1000 \quad & \quad 2000 \quad &\cr
\+ \hfill $N$ \hfill & \hfill 250 \hfill & \hfill 500 \hfill &
\hfill 1000 \hfill & \hfill 2000 \hfill &\cr
\smallskip
\+ \hfill $t_1$ \hfill & \hfill  47 \hfill & \hfill 119 \hfill &
\hfill 262 \hfill & \hfill 500 \hfill &\cr
\+ \hfill $t_3$ \hfill &  \hfill 81 \hfill & \hfill 135 \hfill &
\hfill 284 \hfill & \hfill 550 \hfill &\cr
\+ \hfill $t_{10}$ \hfill & \hfill 87 \hfill & \hfill 151 \hfill &
\hfill 300  \hfill&  \hfill 555 \hfill &\cr
\+ \hfill $t_{coll}$ \hfill & \hfill 95 \hfill & \hfill 164 \hfill &
\hfill 332 \hfill&  \hfill 600 \hfill &\cr}$$
\smallskip
Note: $t_1$ is the median time (over a number of simulations) at which a
binary with energy exceeding $1kT$ first appears.  The quantities $t_3$
and $t_{10}$ are similarly defined for energies of $3kT$ and $10kT$
respectively. The quantity $t_{coll}$ is an estimate of the mean time of
the end of core collapse (cf. \S5.1).

\bigskip
\bigskip
\centerline{\bf Table 3}
\medskip
\centerline{MASS LOSS AND RELAXATION PARAMETERS}
\medskip

$$\vbox{ \settabs \+ ~~1000~~ & ~~0.0820~~ & ~~$t_{min}$~~ & ~~$t_{max}$~~ &
{}~~19.381~~ & ~~0.0543~~ & ~~0.1424~~ & ~~0.0354~~ & ~~0.1598~~ &\cr
\+ \hfill $N$ \hfill & \hfill $\nu$ \hfill & \hfill $t_{min}$ \hfill &
\hfill $t_{max}$ \hfill & \hfill $t_0$ \hfill & \hfill $\gamma_1$ \hfill &
\hfill $C_1$ \hfill & \hfill $\gamma_2$ \hfill & \hfill $C_2$ \hfill &\cr
\smallskip
\+ \hfill 250 \hfill & \hfill  0.083 \hfill & \hfill 360 \hfill &
\hfill 1000 \hfill & \hfill -8.3 \hfill &\hfill 0.054 \hfill &
\hfill 0.142 \hfill &\hfill 0.041 \hfill &\hfill 0.160 \hfill &\cr
\+ \hfill 500 \hfill &  \hfill 0.089 \hfill & \hfill 400 \hfill &
\hfill 1000 \hfill & \hfill -9.1 \hfill &\hfill 0.040 \hfill &
\hfill 0.143 \hfill &\hfill 0.030 \hfill &\hfill 0.158 \hfill &\cr
\+ \hfill 1000 \hfill & \hfill 0.082 \hfill & \hfill 420 \hfill &
\hfill 600  \hfill&  \hfill 19.4 \hfill &\hfill 0.045 \hfill &
\hfill 0.142 \hfill &\hfill 0.035 \hfill &\hfill 0.152 \hfill &\cr
\+ \hfill 2000 \hfill & \hfill 0.090 \hfill & \hfill 700 \hfill &
\hfill 2300 \hfill&  \hfill -45.6 \hfill &\hfill 0.067 \hfill &
\hfill 0.147 \hfill &\hfill 0.044 \hfill &\hfill 0.162 \hfill &\cr}$$
\bigskip
Notes: The parameter $\nu$ is defined in the discussion of eqs.(14) and
(15).  This and other parameters listed have been determined over the
range of times $(t_{min}, t_{max})$.  The parameter $t_0$ is the origin
of time which permits the best fit of eqs.(14) and (15) to the $N$-body
data.  The parameters $\gamma$ and $C$ are, respectively, the
coefficient of $N$ in the argument of the Coulomb logarithm, and a
coefficient in the expression for the conductivity in the gas model.
Subscripts $1$ and $2$ are explained in \S3.1.
\bigskip
\vfill\eject

\centerline{\bf Table 4}
\medskip
\centerline{CONDITIONS AT CORE BOUNCE}
\medskip

$$\vbox{ \settabs \+\quad 2000 \quad & \quad 626 \quad & \quad 218.2 \quad &
\quad 1.110 \quad & \quad 0.034 \quad & \quad 44.7 \quad &\cr
\+ \hfill $N$ \hfill & \hfill $t_{coll}$ \hfill & \hfill $\rho_c$ \hfill &
\hfill $v_c$ \hfill & \hfill $r_c$ \hfill & \hfill $N_c$ \hfill &\cr
\smallskip
\+ \hfill 250 \hfill & \hfill  95 \hfill & \hfill 22.5 \hfill &
\hfill 0.93 \hfill & \hfill 0.098 \hfill & \hfill 12.6 \hfill &\cr
\+ \hfill 500 \hfill &  \hfill 164 \hfill & \hfill 42.3 \hfill &
\hfill 0.97 \hfill & \hfill 0.069 \hfill & \hfill 20.0 \hfill &\cr
\+ \hfill 1000 \hfill & \hfill 332 \hfill & \hfill 125.9 \hfill &
\hfill 1.04  \hfill&  \hfill 0.047 \hfill & \hfill 28.2 \hfill &\cr
\+ \hfill 2000 \hfill & \hfill 600 \hfill & \hfill 281.2 \hfill &
\hfill 1.10 \hfill&  \hfill 0.034 \hfill & \hfill 44.7 \hfill &\cr}$$
\bigskip
Note:  $\rho_c$, $v_c$, $r_c$ and $N_c$ are, respectively, the central
density, the rms three-dimensional speed of stars in the core, the core
radius, and the number of stars within this radius, all measured at the
time of core bounce, $t_{coll}$.
\bigskip
\vfill\eject

{\bf FIGURE CAPTIONS}
\medskip

{\bf Fig.1}  Number of regularised binaries. The time units for $N
= 250$ , $500$ and $2000$ have been scaled to that for $N = 1000$ according
to results of Paper I.  Only binaries which are not escapers (\S6.3) are
included.

{\bf Fig.2}  Maximum binding energy of bound regularised binaries. The time
units have been scaled as in Fig.1.  Recall that the initial binding energy
of each system is $-1/4$.

{\bf Fig.3}  Total ``external" binding energy of all bound members. The time
units have been scaled as in Fig.1.

{\bf Fig 4}  Lagrangian and core radii for $N = 500$.

{\bf Fig.5}  Evolution of the inner Lagrangian radii for one $250$-body model
(case 25). The most energetic interactions of binaries are marked at the lower
border, thick lines indicating binary escapers.

{\bf Fig.6}  Logarithmic gradient profiles of density at several
times (identified in the key).
The radii are scaled by the half mass radius. a) $N$-body models for $N = 250$,
b) $N$-body models for $N = 2000$, c) isotropic gas model for $N = 250$,
d) isotropic gas model for $N = 2000$.  For the gas models $C_b$ is the
value of a coefficient in the binary heating formula, eq.(13), and $t_{b0}$
is the time at which this is switched on (cf. Giersz \& Spurzem 1993).
For the $N$-body models data on Lagrangian radii have first been
smoothed over an interval of order (a) 20 and (b) 70.

{\bf Fig.7} Evolution of the three-dimensional velocity dispersion
with time (a) between the Lagrangian radii corresponding to $1$ and
$2\%$ of the mass, and (b) between the Lagrangian radii corresponding
to $50$ and $75\%$ of the mass, for models with $N = 500$.  The dashed
line gives the corresponding result for an isotropic gas model with
parameters for energy generation given by $C_b = 70$ and $t_{b0} =
130$ (cf. the caption to Fig.6).

{\bf Fig.8}  Velocity dispersion profiles at several times throughout
the pre- and
post-collapse evolution of systems with $N=500$.  The key identifies the
$N$-body time corresponding to each curve.  The radius is scaled by
$r_h$, and the velocity dispersion by the central value.

{\bf Fig.9}  Central potential (potential at the position of the density
centre) for models with $N = 250$, $500$, $1000$ and $2000$. The time units
have been scaled as in Fig.1.

{\bf Fig.10} Central number density for models with $N= 250$, $500$, $1000$
and $2000$.  The time units have been scaled as in Fig.1.

{\bf Fig.11}  Central root-mean-square three-dimensional velocity for
models with $N= 250$, $500$, $1000$ and $2000$.  The times have been scaled
as in Fig.1.

{\bf Fig.12}  Ratio between core and half-mass radii for $N= 250$, $500$,
$1000$ and $2000$.  The times have been scaled as in Fig.1.

{\bf Fig.13}  Number of stars in the core (i.e. the number within core
radius $r_c$) for $N= 250$, $500$, $1000$
and $2000$.  The times have been scaled as in Fig.1.

{\bf Fig.14} Evolution of core parameters $\rho_c$ and $v_c^2$ for $N =
500$ and (dashed curve) for an isotropic gas model specified as in
Fig.7.  The $N$-body data were smoothed using the standard routine SMOOFT
from Press \et (1986).

{\bf Fig.15}  Number of escaping stars for models with $N = 250$, $500$,
$1000$ and $2000$.

{\bf Fig.16}  Dependence of ``external" energy of escapers on time for
$N = 250$, $500$, $1000$ and $2000$. Spikes observed for $N = 250$ are
connected with temporary loss of data on individual models due to hardware
problems.

{\bf Fig.17} Scatter diagram showing the external energies (scaled by
the central value of $kT$) of all single escapers in models with $N =
1000$. Very similar results are obtained for other values of $N$.

{\bf Fig.18}  Number of escaping binaries for models with $N = 250$, $500$,
$1000$ and $2000$.

{\bf Fig.19}  Internal energy of escaping binaries for models with
$N = 250$, $500$, $1000$ and $2000$.

{\bf Fig.20}  Central potential in units of the central one-dimensional
velocity dispersion.  The data have been smoothed as in Fig.14, and
the times have been scaled as in Fig.1. For a Plummer model ($t = 0$)
the theoretical value is $3\phi_c/v_c^2=6$.

{\bf Fig.21} Histograms of the last relative energy change, $\Delta$, for
escaping binaries. Thick line - data directly obtained from $N$-body models;
thin line - data estimated from eq.(20).  The total number of binaries
in each category of data is $N_t$, and the ordinate is the fraction
of cases falling within the corresponding bin.

\bye